\documentclass[twocolumn,showpacs,aps,prl]{revtex4}
\usepackage{graphicx}
\usepackage{dcolumn}
\usepackage{amsmath}
\usepackage{latexsym}

\begin {document}

\title
{
A new route to Explosive Percolation
}
\author
{
S. S. Manna$^{1,2}$ and Arnab Chatterjee$^3$
}
\affiliation
{
\begin {tabular}{c}
$^1$Max-Planck-Institute f\"ur Physik Komplexer Systeme,
    N\"othnitzer Str. 38, D-01187 Dresden, Germany \\
$^2$Satyendra Nath Bose National Centre for Basic Sciences,
Block-JD, Sector-III, Salt Lake, Kolkata-700098, India \\
$^3$CMSPS, The Abdus Salam International Centre for Theoretical Physics,
Strada Costiera 11, Trieste I-34014, Italy.
\end{tabular}
}
\begin{abstract}

      The biased link occupation rule in the Achlioptas process (AP) discourages the large clusters to grow much ahead of 
   others and encourages faster growth of clusters which lag behind. In this paper we propose a model where this tendency
   is sharply reflected in the Gamma distribution of the cluster sizes, unlike the power law distribution in AP. In this 
   model single edges between pairs of clusters of sizes $s_i$ and $s_j$ are occupied with a probability $\propto 
   (s_is_j)^{\alpha}$. The parameter $\alpha$ is continuously tunable over the entire real axis. Numerical studies indicate 
   that for $\alpha < \alpha_c$ the transition is first order, $\alpha_c=0$ for square lattice and $\alpha_c=-1/2$ for 
   random graphs. In the limits of $\alpha = -\infty, +\infty$ this model coincides with models well established in the 
   literature.

\end{abstract}
\maketitle

      A first order transition in critical phenomena is characterized by an abrupt jump in the order parameter 
   \cite {Stanley}. An infinitesimal increase of the control variable leads to a spurt of activity on the global 
   scale as in a catastrophic process. The percolation transition describes a geometrical phase transition 
   between ordered and disordered phases of random resistor network, binary alloys, forest fires, galaxies etc.
   which are continuous transitions \cite {Grimmett,Stauffer,Kast}. On the other hand first order transition is observed in
   Bootstrap Percolation \cite {Chalupa} modeling the competition between exchange and crystal field 
   interaction in some magnetic materials. Here, depending on the lattice structure, it may happen that culling 
   of even a single spin evacuates a globally connected system in a recursive process.

      Recently it is observed that a biased link occupation rule, known as the Achlioptas process \cite {ACH}, 
   on the Erd\H os-R\'enyi random graph \cite {Erdos} leads to a first order transition. Here a pair of edges $(ij)$ and $(kl)$ 
   are randomly selected between clusters of sizes $s_i$, $s_j$ and $s_k$, $s_l$ respectively. The edge with the 
   smaller value of the products $s_is_j$ and $s_ks_l$ is occupied; when equal, one edge is selected randomly.
   Repeated application of this rule for every edge delays growth of the largest cluster. Therefore the biased edge 
   occupation rule in AP discourages a single cluster to grow much ahead of other clusters and at the same time it 
   encourages faster growth of clusters which lag behind. One therefore expects that at a certain stage the cluster 
   size distribution should have a characteristic scale. Finally nearly equal size clusters are linked together by additional few 
   edges (of zero measure) resulting an abrupt global connection as in first order transition, hence the name `Explosive Percolation' 
   (EP) \cite {ACH}, triggering an explosion of scientific activity. Later, details of AP on square lattice 
   \cite {ZIFF} and random graph \cite {Friedman} have been studied. Also AP on scale-free networks observes a transition 
   from a continuous to a first order transition \cite {Cho,Radicchi,Radicchi1}. A Hamiltonian formulation of AP is studied in \cite {Herrmann}. 
   first order transition has been observed in a cluster aggregation process \cite {Cho1} and also in human protein homology network 
   \cite {Rozenfeld}. However, very recently, Costa et. al. has claimed that the nature of transition in Explosive Percolation
   is indeed continuous \cite {Dorogov}. Opposite bias can also be given by occupying the edge with larger product and this
   percolation transition remains continuous.

\begin{figure}[top]
\begin{center}
\includegraphics[width=3.5cm]{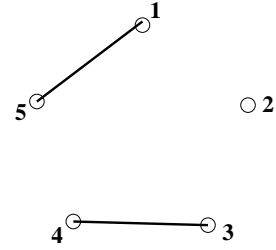}
\end{center}
\caption{
A small graph of 5 nodes and 2 links. The next link is placed using the probabilty $\pi_{ij}$
as defined in the text.
}
\end{figure}

      In this paper we argue that it is possible to attain EP in a generic fashion by introducing a continuously 
   tunable parameter and selecting only a single edge at a time. At each step in the process we consider all 
   currently vacant edges (if we wish to avoid edges internal to existing clusters, the loop-less condition defined below, 
   we consider only those edges which connect clusters). Let the sizes of the respective clusters to be connected by the edges be 
   denoted $s_i$ and $s_j$. Each of these vacant edges is given a weight $(s_is_j)^{\alpha}$ and the sum over all 
   the weights is calculated as a normalization constant $W$. Then one of these vacant edges is chosen with probability 
   $(s_is_j)^\alpha/W$ and occupied. This procedure is then repeated for the next edge addition. 

\begin{figure*}[top]
\begin{center}
\includegraphics[width=10.5cm]{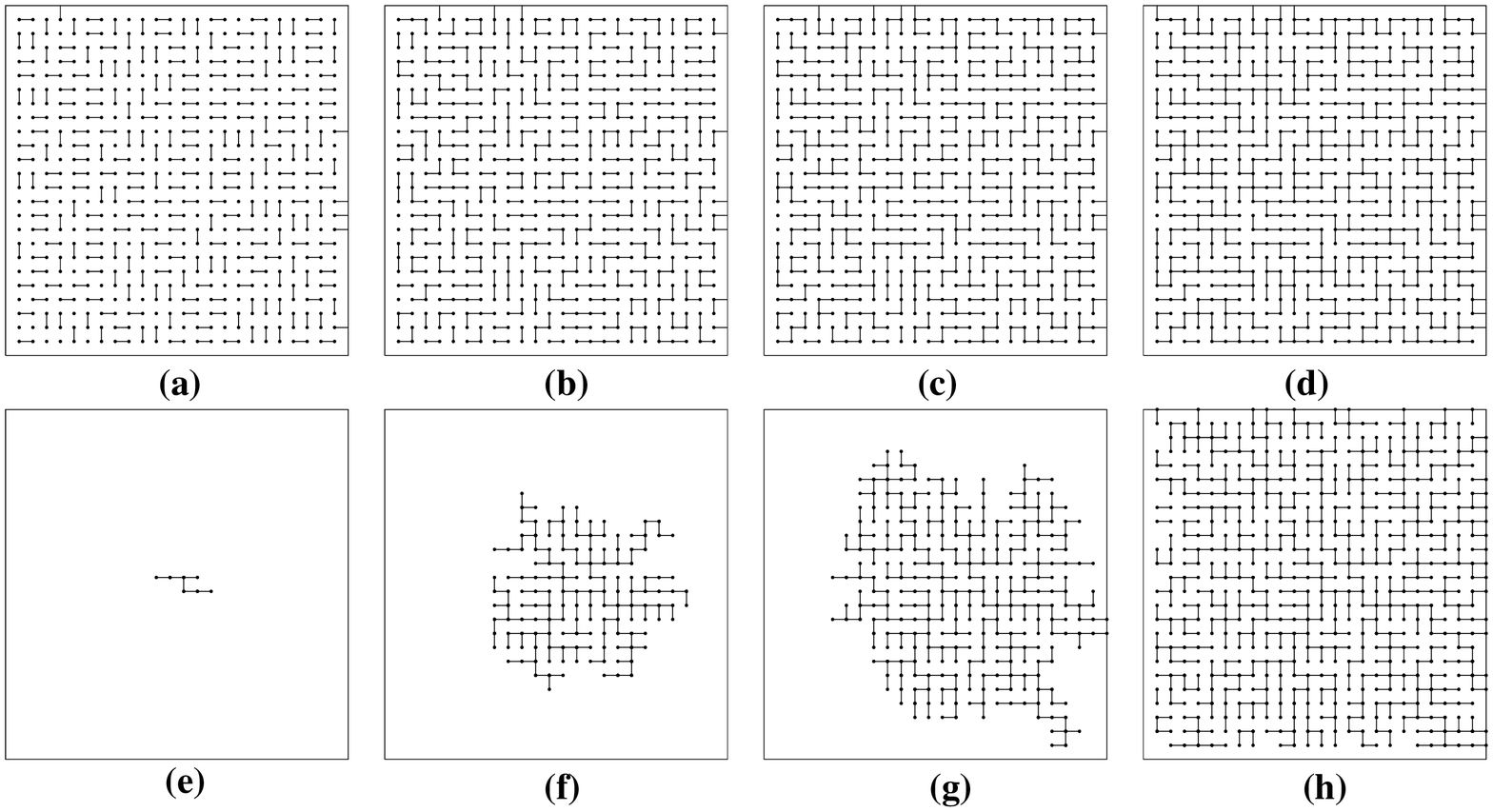}
\end{center}
\caption{
Percolation with $\alpha \to -\infty$ on square lattice ($L=24$) and with loop-less condition. (a) Only links are deposited (b) clusters of 
size up to 4 (c) clusters of size up to 8 and the (d) the spanning tree. Similarly percolation with $\alpha \to +\infty$ are:
(e) growth starts from a single link (f) cluster of $N/4$ bonds (g) cluster of $N/2$ bonds (h) the spanning tree
cluster.
}
\end{figure*}

      For example, let us consider a $N$ = 5 nodes graph (Fig. 1). Let at an intermediate stage there be only two links 
   in this graph, connecting nodes 1 and 5 and nodes 3 and 4. Therefore the graph has three components of sizes 2, 2, 1 
   and has 8 vacant edges: (1,2), (1,3), (1,4), (2,3), (2,4), (2,5), (3,5) and (4,5). Now, to place the third link in 
   this graph we calculate weights associated with every remaining vacant edge of the graph. For some specific pre-assigned 
   value of $\alpha$ the associated weights are:
$w_{12}$ = $(2.1)^{\alpha}$,
$w_{13}$ = $(2.2)^{\alpha}$,
$w_{14}$ = $(2.2)^{\alpha}$,
$w_{23}$ = $(1.2)^{\alpha}$,
$w_{24}$ = $(1.2)^{\alpha}$,
$w_{25}$ = $(1.2)^{\alpha}$,
$w_{35}$ = $(2.2)^{\alpha}$,
$w_{45}$ = $(2.2)^{\alpha}$. 
The sum of all these weights: $W = \Sigma w_{ij}$ = $w_{12}$+$w_{13}$+$w_{14}$+$w_{23}$+$w_{24}$+$w_{25}$+$w_{35}$+$w_{45}$
and the edge occupation probablities are $\pi_{ij} = w_{ij}/W$. One vacant edge is then randomly selected with probability 
$\pi_{ij}$ and is occupied. This completes the procedure to occupy one edge which is then repreated.

      Here $\alpha$ being a 
   continuously tunable parameter varying over the entire real axis from $-\infty$ to $+\infty$. This is the crucial 
   difference between AP and our model. In AP both links are picked up with uniform probability as in random graph, i.e., each link
   connects clusters with probability $\propto s_is_j$. In comparison, in our case 
   links are picked up using cluster size dependent probabilities. For $\alpha < 0$, edges with smaller products are 
   preferentially selected which delays the transition and in the following we present numerical evidence that for all 
   $\alpha < \alpha_c$ the percolation transition is first order. For $\alpha > 0$ edges with larger products are occupied with higher 
   probabilities. Our model is completely distinct from AP, i.e., AP cannot be retrieved for any value of $\alpha$ 
   except in two limits. AP can be extended by selecting the minimal product of $n$ randomly selected edges. In the 
   limit of $n \to \infty$ the edge which has the globally minimal (maximal) value of the product is only selected. 
   Only these two limits iof extended AP correspond to $\alpha \to -\infty(+\infty)$ of our model. 

      The `sequential time' $t$ is defined as the number of links occupied and the link densities are $r=t/N$ for 
   random graph and $p=t/(2N)$ for square lattice with periodic boundary condition. We consider percolation with (without) 
   loops where links are allowed (forbidden) to 
   connect two nodes of the same cluster. The ordinary bond percolation corresponds to $\alpha=0$. When `loop-less'
   condition is imposed this is called `loop-less percolation' \cite {Manna1} with $p_c = (7-3\sqrt3)/4$ on square lattice
   \cite {ZIFF}. Since random graph is infinite dimensional loop-less condition has no effect and $r_c = 1/2$.

      Next, the $\alpha \to -\infty$ loop-less case is considered. Here at any stage one edge is randomly occupied only from the 
   subset of all vacant edges for which the $s_is_j$ values are minimum. In the beginning all $N$ nodes are isolated and 
   the minimal product (MP) is 1. The MP condition is essentially a hard core repulsion and therefore successive links are 
   placed without overlapping with the previous ones. This process is known as the ``random sequential adsorption''
   \cite {Manna}. Gradually the system reaches a jamming limit when no more link can be placed (Fig. 2(a)). The jammed 
   state of random graph has $N/2$ links but there are isolated nodes and links in square lattice. At this stage the MP jumps to 4 for random graph but to 
   2 for square lattice. Consequently the next sequence of links connect pairs of links in random graph but single sites with links in square lattice. 
   Eventually the vacant edges with this MP are again exhausted at a second jammed state when the MP is further enhanced 
   and this process continues in a series of jammed states with discrete jumps of MPs (Fig. 2(b-c)). The sequence of MPs 
   for square lattice are 1, 2, 4, 6 etc. with jamming densities 0.227(1), 0.273(1), 0.351(1), 0.370(1) respectively which approach the 
   $p_c=1/2$, a spanning tree configuration (Fig. 2(d)). In contrast the jamming densities in random graph are $(s-1)/s$ for clusters 
   of sizes $s = 2, 4, 8, 16, ...$ etc. giving $r_c=1$. The largest cluster size when plotted with the link density gives 
   a step function at $r=1$ in $N \to \infty$, a perfect first order transition also observed in \cite {Friedman}. Similarly for 
   $\alpha \to +\infty$ vacant edges with maximal values of $s_is_j$ are only occupied (Fig. 2(e-h)). Once the first link is placed, the 
   subsequent links get connections to this cluster only, therefore only one cluster grows. In random graph a new node is connected 
   to a randomly selected node of the growing cluster. This is the `model A' network of \cite {Barabasi} with an exponentially 
   decaying degree distribution. On square lattice the growth is limited to only the surface bonds leading to the well known 
   `Bond Eden Tree' \cite {BET}. Without loop-less restriction all vacant edges within the cluster are first 
   occupied and bulk of the cluster grows like a perfect crystal.

      Now we study the percolation process with loops for finite value of $\alpha$. Fig. 3(a) plots the
   order parameters ${\cal C}(p,\alpha)=s_m(p,\alpha)/N$, $s_m$ being the size of the largest cluster, for
   square lattice with size $L$ = 512; $N=L^2$. As $\alpha$ decreases from zero the order parameter increases more 
   rapidly and the whole curve shifts to the larger $p$ values indicating increasingly delayed transitions. For 
   $-5 \le \alpha \le -1$ the curves are nearly vertical, ${\cal C}(p,\alpha)$ jumps from 0 to 1 within a short 
   interval of $p$. However within $-1 < \alpha < 0$, the curves are relatively smooth but gradually become steeper 
   as $N$ increases. In the following we present evidence that for all $\alpha < \alpha_c = 0$ the percolation transitions are likely to be first order transition, 
   with $p_c > 1/2$.

\begin{figure}[top]
\begin{center}
\includegraphics[width=5.0cm]{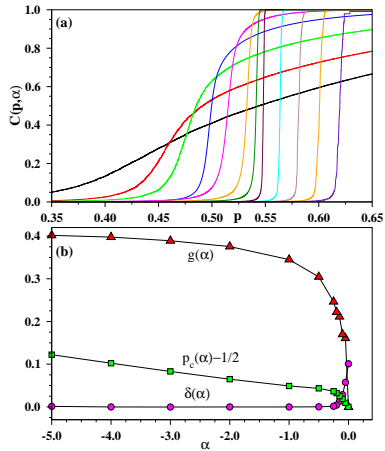}
\end{center}
\caption{(Color online)
(a) Order parameter ${\cal C}(p,\alpha)$ with link density $p$ for $\alpha$ = 1/2, 0.2, 0.1, 0, -0.1, -1/4, 
    -1/2, -1, -2, -3, -4 and -5, $\alpha$ values decreasing from left to right, for square lattice of $L=512$. 
(b) The asymptotic values of the gap $\delta(\alpha)$, the percolation threshold $p_c(\alpha)-1/2$ and the 
    largest jump $g(\alpha)$ of the order parameter plotted with $\alpha$.
}
\end{figure}

\begin{figure}[top]
\begin{center}
\includegraphics[width=5.0cm]{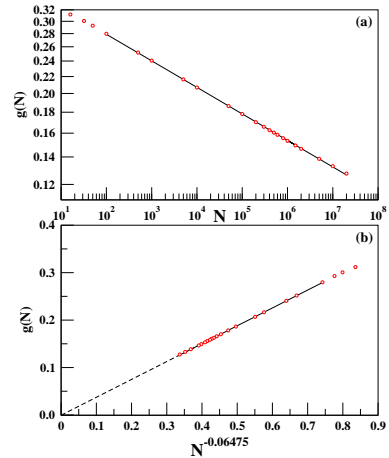}
\end{center}
\caption{(Color online)
(a) For AP on random graph, the maximal jump $g(N)$ in the order parameter has been plotted on a double logarithmic scale with 
    graph size $N$. The slope is -0.06475.
(b) The maximal jump $g(N)$ is then plotted against $N^{-0.06475}$ on a linear scale. The continuous line is a straight line fit 
    of this data which is then extrapolated to $N \to \infty$ to meet $g(N)$ axis at -0.000516.
}
\end{figure}

      To determine the order of transition and estimate the percolation threshold we define: $t_0$ is the latest time with 
   $s_m < N^{1/2}$ and $t_1$ is the earliest time $s_m > \kappa N$; $\kappa=1/2$ (random graph) \cite {ACH} and 0.9 (square lattice). A gap 
   $\Delta = t_1-t_0$ in $N \to \infty$ limit can distinguish between two kinds of transitions: 
   $\delta(\alpha)=\lim_{N\to\infty}\frac{\Delta(\alpha,N)}{N}= constant > 0$ 
   for a continuous transition and 0 for a first order transition. Average values of $\Delta(\alpha,N)/N$ are then extrapolated to $N \to \infty$ 
   limit as $N^{-\beta(\alpha)}$. Different $\beta(\alpha)$ values for an $\alpha$ are tried and the one with the minimal 
   Standard Error \cite {Wolfram} is selected. The $\delta(\alpha)$ values are nearly zero for $-5 \le \alpha \le -1/4$ indicating 
   the first order transition (Fig. 3(b)). There is a gradual increase of $\delta(\alpha)$ within $-1/4 < \alpha < 0$ due to the finite size of 
   systems simulated. However $\delta(0) \approx 0.101$ is obtained for ordinary percolation. 

      We also studied in detail the extent of jump in the order at the transition point. For that we calculated the incremental 
   changes in ${\cal C}$ due to the occupation of an additional link and keep track of these changes as links are occupied one by one.
   The average value of the largest jump $\Delta {\cal C}_m$ due to addition of only a single link and the corresponding link 
   density $r_m$ have been calculated. In the thermodynamic limit of $N\to \infty$ the asymptotic jump length is:
   $g= \lim_{N\to\infty}g(N)= \lim_{N\to\infty}\Delta{\cal C}_m(N)$. We expect $g > 0$ for a discontinuous transition and
   $g = 0$ for a continuous transition. We first calculate this quantity for the AP on random graph. This data has been shown 
   in Fig. 4 and $g(N)$ is observed to vary as: $g(N) \sim N^{-0.064771}$ which implies that $g(N)$ approaches to $g=0$
   as $N \to \infty$. We believe this result is consistent with the claim in \cite {Dorogov} that the AP is a continuous 
   transition.

      For our problem the asymptotic jump 
   $g(\alpha) = \lim_{N\to\infty}\Delta {\cal C}_m(\alpha,N)$ for different $\alpha$ are plotted in Fig. 3(b). For ordinary bond 
   percolation $g(\alpha=0,N) \to 0$ as $N^{d_f/2-1}$, with $d_f=91/48$, the fractal dimension of the incipient infinite 
   percolation cluster \cite {Stauffer}. However for $\alpha < 0$ the $g(\alpha)$ jumps to 0.16 at $\alpha=-0.05$ and then gradually 
   increases to 1/2 as $\alpha \to -\infty$. The percolation thresholds $p_c(N)$ are estimated by the average values of $r_m$ and $t_1$ 
   giving approximately same asymptotic value for $p_c(\alpha)$. The $p_c(0)=\lim_{N\to\infty}p_c(0,N) \to 1/2$ limit is 
   approached as $N^{-1/2\nu}$, $\nu = 4/3$ \cite {Stauffer} being the correlation length exponent for the ordinary percolation. 
   On the other hand for $\alpha <0$, $p_c(\alpha)$ increased continuously with $\alpha$ (Fig. 3(b)) and tends to unity as $\alpha \to -\infty$.

      In Fig. 5(a) we show ${\cal C}(r,\alpha)$ vs. $r$ for random graph.
   Last five curves for $\alpha \le -1$ almost coincide with the vertical line at $r=1$ implying first order transition 
   with $r_c=1$ in this range. Curves are smoother for $-1 < \alpha < 0$. In Fig. 5(b) we show $r_c(\alpha)$ tends to 1/2 
   rapidly as $\alpha \to 0$. The jump $g(\alpha)$ is almost zero for $-1/4 \le \alpha \le 0$ but then it rapidly increases 
   and tends to 1/2 as $\alpha \to -\infty$. The gap $\delta(\alpha)$ is almost zero for $\alpha \le -1/2$ and then slowly 
   increases to $\approx 0.193$ for random graph. This result indicates that for random graph $\alpha_c=-1/2$.

\begin{figure}[top]
\begin{center}
\includegraphics[width=5.0cm]{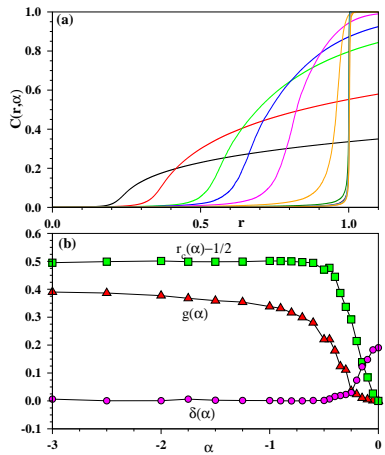}
\end{center}
\caption{(Color online)
(a) Order parameter ${\cal C}(r,\alpha)$ with link density $r$ for $\alpha$ = 1/2, 1/4, 0, -0.1, -1/4, 
    -1/2, -1, -2, -3, -4 and -5, $\alpha$ values decreasing from left to right, for random graph of $N=4096$. 
(b) The asymptotic values of the gap $\delta(\alpha)$, the percolation threshold $r_c(\alpha)-1/2$ and the 
    largest jump $g(\alpha)$ of the order parameter plotted with $\alpha$.
}
\end{figure}

      The approach to first order transition in our system is completely different from AP. Comparison is made in the tendency of supressing 
   growth of large clusters and enhancing growth of small clusters as reflected in following three quantities. In Fig. 6(a) 
   we show the scaled cluster size distribution $P(s,p)$ for square lattice at $\alpha=-1$. The individual distributions has strong 
   dependence on $\Delta p=p_c-p$ but not significantly on $L$. The best scaling form is $P(s,p)\Delta p^{-\eta} \sim 
   {\cal G}(s \Delta p^{\zeta})$ with $\eta=\zeta=1.77(5)$. The scaling function fits to the Gamma distribution ${\cal G}(x) 
   \sim x^{a}\exp(-bx)$ with $a = 0.93(10)$ and $b = 4.09(10)$. $P(s,p)$ for $\alpha < -1$ also fits to Gamma distributions 
   with different values of $a$ and $b$. In addition we have checked the finite size scaling analysis as well. Right at the 
   percolation point the cluster size distribution $P(s,P_c,L)L^{0.9}$ scales excellent with $s/L^{0.9}$ and the scaling function fits
   very well to a Gamma distribution. Here the percolation point is determined by the maximal jump in the order parameter.

      In comparison $P(s,r)$ in AP follows a power law as in continuous transition and 
   the exponent approaches to $\approx -2.11(2)$ as $r \to r_c$. As $r_c$ is approached, a hump appears in the tail 
   implying an enhanced population of large clusters. At percolation point these clusters get linked giving rise to a sudden 
   jump of the size of the largest cluster. We believe this is due to random selection of edges in AP the power law distribution 
   masks the Gamma distribution; the exponent is raised to -2.11 from -5/2 in random graph but retains -2.06(1) for square lattice 
   (Fig. 6(b)) compared to 187/91 for ordinary percolation in 2$d$ also observed in \cite {Radicchi1}.

\begin{figure}[top]
\begin{center}
\includegraphics[width=5.0cm]{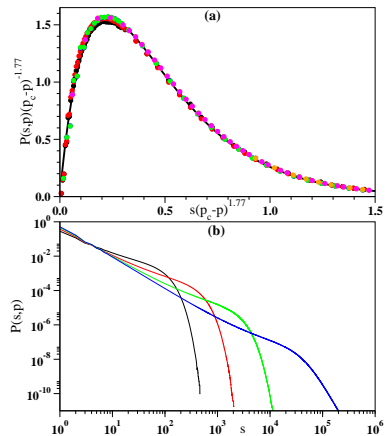}
\end{center}
\caption{(Color online)
(a) Finite-size scaling of the cluster size distribution $P(s,p)$ of square lattice at $\alpha=-1$ (with $p_c=0.549(2)$) for $L$= 64 
    and 128; $p$ = 0.48, 0.50 and 0.52. The solid line is a fit to the Gamma distribution.
(b) The binned average cluster size distribution $P(s,p)$ of AP on square lattice for $L=1024$; $p$ = 0.49, 0.51, 0.52 and 0.525
    (from left to right).
}
\end{figure}

      Since in first order transition there are many clusters whose sizes are nearly same we calculate the average ratio ${\cal R}(p) = 
   \langle s_m/s_{nm} \rangle$ where $s_{nm}$ is the size of the second largest cluster.  For first order transition we expect ${\cal R}(p)$ 
   to maintain a uniform variation with $p$ where as it should increase smoothly beyond the percolation threshold for a 
   continuous transition. Fig. 7(a) shows ${\cal R}(p)$ for square lattice with $L=128$ and AP on square lattice. For all $\alpha < 0$ the plots 
   are indeed horizontal 
   lines up to $p = p_c(\alpha)$. It implies that $s_m$ and $s_{nm}$ have comparable values till the system reaches $p_c$
   and the lines stop shortly after $p_c$ since at percolation threshold the largest cluster covers the entire system. In 
   comparison ${\cal R}(p)$ for $\alpha=0$ and AP on square lattice grows continuously to very large values indicating that the largest cluster grows 
   by swallowing small clusters and it continues far beyond $p_c$. A third quantity is the decay of the fraction $n_1(p,\alpha)$ 
   of isolated nodes in the system. For the first order transition, the $n_1(p,\alpha)$ decays very fast and vanishes at or 
   before $p_c$ for all $\alpha < 0$ (Fig. 7(b)). It is also observed that as $\alpha$ decreases 
   $n_1(p,\alpha)$ vanishes even faster, e.g., as $\alpha \to -\infty$, $n_1(p,-\infty)$ decreases 
   linearly to zero at $p=0.273$. In comparison $n_1(p)$ decays as $\exp(-2r)$ in random graph and $\exp(-r^{1.17})$ in the AP,
   though on square lattice they behave different from our model (Fig. 7(b)).
   We argue that the cluster size distribution $P(s,p)$, ratio ${\cal R}(p)$ of two largest clusters and the density of 
   isolated nodes $n_1(p)$ reflect distinguishing signatures of our model with AP at, beyond and before the transition.

      Finally we like to mention that numerical study of our model is CPU intensive since for $\alpha \ne 0$ the CPU $\sim N^3$ for
   random graph and $\sim N^2$ for square lattice in comparison to $\sim N$ in AP. Therefore we could study up to $L = 512$ for square lattice and $N=4096$
   for random graph, though the extent of our computational efforts are comparable to other studies in the literature.

\begin{figure}[top]
\begin{center}
\includegraphics[width=5.0cm]{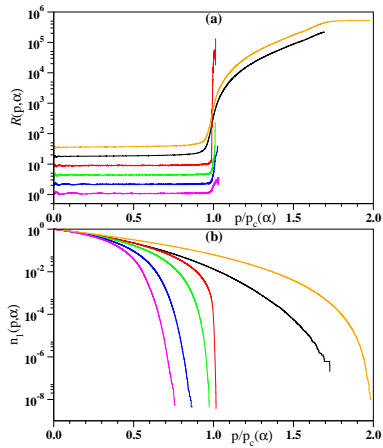}
\end{center}
\caption{(Color online)
(a) Ratio ${\cal R}(p,\alpha)$ of the sizes of largest cluster and the next largest with $p/p_c(\alpha)$ for $\alpha=-3,-2,-1,-1/2, 0$ 
    and AP on square lattice from bottom to top. For more visibility $y$ axes have been multiplied by 1, 2, 4, 8, 16
    and 32 respectively.
(b) Fraction $n_1(p,\alpha)$ of isolated nodes with $p/p_c(\alpha)$ for the same $\alpha$ values (from right to left)
    as in (a).
}
\end{figure}

      To summarize, we have studied a new route to Explosive Percolation where single edges between pairs of clusters are randomly 
   occupied with probabilities $\propto (s_is_j)^{\alpha}$. From our numerical studies we present evidence that for 
   all $\alpha < \alpha_c$ the transitions are discontinuous first order transitions. We obtain $\alpha_c=0$ for the
   square lattice but -1/2 for random graphs. Our model is completely
   different from AP apart from the limits of $\alpha \to -\infty$ and $+\infty$ where it coincides with similar limits of the
   generalized AP. The effect of the biased occupation rule discouraging growth of large clusters and encouraging growth of small
   clusters is distinctly visible in the Gamma distribution of the cluster sizes in our model. This is in contrast to power law
   distribution in AP like the continuous transition which we believe is due to random selection of edges. In addition 
   we define two quantities ${\cal R}(p)$ and $n_1(p)$ which distinguishes our model from AP. 

\leftline {manna@bose.res.in}

\end {document}